\documentstyle [12pt]{article}
\def\int {\intop \limits}

\def\fnote#1{\footnote}
\begin{document}
\newcommand{\dst}[1]{\displaystyle{#1}}
\newcommand{\barl}{\begin{array}{rl}}
\newcommand{\ball}{\begin{array}{ll}}
\newcommand{\ear}{\end{array}}
\newcommand{\barc}{\begin{array}{c}}
\newcommand{\sne}[1]{\displaystyle{\sum _{#1} }}
\newcommand{\sn}[1]{\displaystyle{\sum ^{\infty }_{#1} }}
\newcommand{\ini}[1]{\displaystyle{\int ^{\infty }_{#1}}}
\newcommand{\myi}[2]{\displaystyle{\int ^{#1}_{#2}}}
\newcommand{\inn}{\displaystyle{\int }}
\newcommand{\be}{\begin{equation}}
\newcommand{\ee}{\end{equation}}
\newcommand{\aq}[1]{\label{#1}}
\renewcommand \theequation{\thesection.\arabic{equation}}

\vspace*{4.0cm}
\centerline{\Large {\bf The Landau-Pomeranchuk-Migdal effect}}
\vskip .25cm
\centerline{\Large {\bf in a thin target}}
\vskip .5cm
\centerline{\large{\bf V. N. Baier and V. M. Katkov}}
\centerline{Budker Institute of Nuclear Physics,
 630090 Novosibirsk, Russia}
\vskip 2.0cm
\begin{abstract}
The Landau, Pomeranchuk, Migdal (LPM) effect (suppression of the
bremsstrahlung from high energy electron
due to a multiple scattering of an emitting electron in dense media)
is considered for the case when thickness of a target is of the order
or less than the formation length of radiation.
The effects of the polarization of a medium
and transition radiation are taken into account as well.
Qualitative picture of the phenomenon is discussed in detail.
Comparison with recent experimental data is carried out.
\end{abstract}

\newpage
\section{Introduction}

The process of bremsstrahlung from high-energy electron
occurs over a rather long distance,
known as the formation length. If anything happens to an electron
or a photon while traveling this distance, the emission can be disrupted.
Landau and Pomeranchuk showed that if the formation
length of bremsstrahlung becomes comparable to the distance over which
a mean angle of multiple scattering becomes comparable with a
characteristic angle of radiation, the bremsstrahlung will be
suppressed \cite{1}. Migdal \cite{2}, \cite{3} developed a quantitative
theory of this phenomenon.
An influence of polarization of a medium on radiation process
leads also to suppression of the soft photon emission (Ter-Mikaelian
effect, see in \cite{4}).

A very successful series of experiments \cite{5} - \cite{7} was performed
at SLAC during last years. In these experiments the cross section
of bremsstrahlung of soft photons with energy from 200~KeV to
500~MeV from electrons with energy 8~GeV and 25~GeV is measured
with an accuracy of the order of a few percent. Both LPM and dielectric
suppression is observed and investigated. These experiments were the
challenge for the theory since in all the previous papers calculations
(cited in \cite{8}) are performed to
logarithmic accuracy which is not enough for description
of the new experiment. The contribution of the Coulomb corrections (at least
for heavy elements) is larger then experimental errors and these corrections
should be taken into account.

Very recently authors developed the new approach to the theory of LPM
effect \cite{8}
where the cross section of bremsstrahlung process
in the photon energies region where the influence of the LPN is very strong
was calculated with term $\propto 1/L$ , where $L$
is characteristic logarithm of the problem,
and with the Coulomb corrections
taken into account. In the photon energy region, where the LPM effect
is "turned off", the obtained cross section
gives the exact Bethe-Heitler cross section (within power accuracy) with
Coulomb corrections. This important feature was absent in
the previous calculations. The contribution of an inelastic scattering of
a projectile on atomic electrons is also included.
The polarization of a medium is incorporated into
this approach. The considerable contribution into the soft part of
the measured spectrum of radiation gives a photon
emission on the boundaries of a target. We calculated this contribution
taking into account the multiple scattering and polarization of a medium
for the case when a target is much thicker than the
formation length of the radiation. We considered also
a case when a target is much thinner than the
formation length. A case of an intermediate thickness of a target (between
cases of a thick and a thin target) is analyzed but
polarization of a medium is not included.

In the present paper we calculated the cross
section of bremsstrahlung process in a target of intermediate thickness.
In Section 2 we derived general expression for the spectral
probability of radiation in a thin target and in a target with
intermediate thickness where the multiple scattering, the
polarization of a medium and radiation on boundaries of a target
are taken into account. The representations suitable for numerical
calculations are derived. Useful asymptotic formulae are found.
In Section 3 qualitative picture of the phenomenon is discussed in detail.
In Section 4 we compare the calculated spectral curves with recent
experimental data \cite{7} where electrons with energy $\varepsilon=25~GeV$
and $\varepsilon=8~GeV$ radiated in a gold target with thickness
$l=0.7(0.1)\%~L_{rad}$. Agreement between theory and data is perfect
for $l=0.7\%~L_{rad}$ at electron energy $\varepsilon=25~GeV$, for
the same target and $\varepsilon=8~GeV$ agreement is satisfactory.

\section{Spectral distribution of the probability of radiation}
\setcounter{equation}{0}

Proceeding from the formulation of \cite{8} (see Section 4)
we can obtain general expression
which takes into account boundary effects for a target of arbitrary
thickness. With allowance for multiple
scattering and polarization of a medium we
have for the spectral distribution of the probability of radiation
\begin{equation}
\begin{array}{ll}
\displaystyle{\frac{dw}{d\omega}= \frac{4\alpha}{\omega}{\rm Re}
\int_{-\infty}^{\infty}dt_2
\int_{-\infty}^{t_2}dt_1 \exp \left(-i\int_{t_1}^{t_2}\mu(t)dt \right)}\\
\displaystyle{\times \left<0|r_1 S(t_2,t_1)+r_2 {\bf p}
S(t_2, t_1) {\bf p}|0 \right>},
\end{array}
\label{1}\end{equation}
where
\begin{equation}
\begin{array}{ll}
\displaystyle{\mu(t)=\vartheta(-t)+\vartheta(t-T)+\kappa \vartheta(t)
\vartheta(T-t),\quad T=\frac{l}{l_0},\quad l_0=\frac
{2\varepsilon \varepsilon'}{\omega m^2}},\\
\displaystyle{r_1 = \frac{\omega^2}{\varepsilon^2},\quad
r_2=1+\frac{\varepsilon'^2}{\varepsilon^2},\quad \kappa=1+\kappa_0^2,\quad
\kappa_0=\frac{\omega_p}{\omega}},\\
\displaystyle{\omega_p=\gamma \omega_0,\quad
\gamma=\frac{\varepsilon}{m}, \quad \omega_0^2=\frac{4\pi \alpha n}{m}},
\end{array}
\label{2}\end{equation}
here $\varepsilon$ is the energy of the initial electron, $\omega$
is the energy of radiated photon, $\varepsilon'=\varepsilon-\omega$,
$n$ is the density of the atoms in a medium,
$l$ is the thickness of a target.
So, we split time interval
(in the used units) into three parts: before target ($t<0$), after target
($t>T$) and inside target ($0 \leq t \leq T$).
The mean value in Eq.(\ref{1}) is taken
over states with definite value of the two-dimensional operator
$\mbox{\boldmath$\varrho$}$ (see \cite{8}, Section 2). The propagator
of electron has a form
\begin{equation}
S(t_2, t_1)={\rm T} \exp
\left[-i\int_{t_1}^{t_2} {\cal H}(t) dt \right],
\label{3}\end{equation}
where the Hamiltonian ${\cal H}(t)$ in the case of a homogeneous
medium is
\begin{equation}
\begin{array}{ll}
\displaystyle{{\cal H} (t)={\bf p}^2-iV(\mbox{\boldmath$\varrho$})g(t),
\quad {\bf p}=-i\mbox{\boldmath$\nabla$}_{\mbox{\boldmath$\varrho$}},
\quad g(t)=\vartheta(t)\vartheta(T-t)},\\
\displaystyle{V(\mbox{\boldmath$\varrho$})=Q\mbox{\boldmath$\varrho$}^2
\left(L_1+\ln \frac{4}{\mbox{\boldmath$\varrho$}^2}-2C \right), \quad
Q=\frac{2\pi Z^2\alpha^2\varepsilon \varepsilon' n}{m^4\omega},\quad
L_1=\ln \frac{a_{s2}^2}{\lambda_c^2},}\\
\displaystyle{\frac{a_{s2}}{\lambda_c}=183Z^{-1/3}{\rm e}^{-f},\quad
f=f(Z\alpha)=(Z\alpha)^2\sum_{k=1}^{\infty}\frac{1}{k(k^2+(Z\alpha)^2)},\quad
C=0.577... .}
\end{array}
\label{4}\end{equation}
The contribution of scattering of a projectile on atomic electrons may
be incorporated into effective potential $V(\mbox{\boldmath$\varrho$})$.
The summary potential including both an elastic and an inelastic scattering
is
\begin{equation}
V(\mbox{\boldmath$\varrho$})+V_e(\mbox{\boldmath$\varrho$})
=-Q_{ef}\mbox{\boldmath$\varrho$}^2
\Big(\ln \gamma^2 \vartheta_{ef}^2
+\ln \frac{\mbox{\boldmath$\varrho$}^2}{4}+2C \Big),
\label{4a}\end{equation}
where
\[
\displaystyle{Q_{ef}=Q(1+\frac{1}{Z}),\quad \vartheta_{ef}
= \vartheta_1 \exp \left[\frac{1}{1+Z}\left(Zf(\alpha Z)-1.88
\right)-\frac{1}{2} \right]}.
\]

In (\ref{1}) it is implied that subtraction is made at
$V=0$,~$\kappa=1$.

It is important to note that Eq.(4.1) of Ref.8 is valid for description
of effects of multiple scattering and polarization of a medium.
But for description of the transition radiation on two boundaries
it should be modified as it is done in Eq.(\ref{1}).

In \cite{8} the potential $V(\mbox{\boldmath$\varrho$})$
was presented in the form
\begin{equation}
\begin{array}{ll}
V(\mbox{\boldmath$\varrho$})=V_c(\mbox{\boldmath$\varrho$})
+v(\mbox{\boldmath$\varrho$}),\quad V_c(\mbox{\boldmath$\varrho$})=
q\mbox{\boldmath$\varrho$}^2, \quad q=QL,\\
\displaystyle{L \equiv L(\varrho_c)=\ln \frac{a_{s2}^2}{\lambda_c^2 \varrho_c^2},\quad
v(\mbox{\boldmath$\varrho$})=-\frac{q\mbox{\boldmath$\varrho$}^2}{L}
\left(\ln \frac{\mbox{\boldmath$\varrho$}^2}{4\varrho_c^2}+2C \right)},
\end{array}
\label{5}\end{equation}
where the parameter $\varrho_c$ is defined by a set of equations:
\begin{equation}
\varrho_c=1~{\rm for}~\nu_1 \leq 1;\quad
4Q\varrho_c^4L(\varrho_c)=1~{\rm for}~\nu_1 \geq 1;\quad
\nu_1^2 \equiv 4QL_1.
\label{6}\end{equation}
This form is convenient for expansion over powers of $1/L$ (typical
value $L \approx 10$).

The formation length of radiation (for $\omega \ll \varepsilon$)
inside target with regard
for the multiple scattering and the polarization of a medium
(see Eqs.(3.3), (7.1) in \cite{8})
\begin{equation}
l_f=\frac{2\gamma^2}{\omega}\left[1 + \gamma^2\vartheta_c^2
+\left(\frac{\gamma \omega_0}{\omega} \right)^2 \right]^{-1}
\label{7}\end{equation}
can be written as
\begin{equation}
\begin{array}{ll}
\displaystyle{l_f=\frac{l_0}{\nu_0+\kappa},\quad \frac{l}{l_f}=T(\nu_0+\kappa),\quad
T=\frac{lm^2\omega}{2\varepsilon \varepsilon'}};\\
\nu_0^2(\varrho_c)=4QL(\varrho_c), \quad \nu_0^2(1)=\nu_1^2
\end{array}
\label{8}\end{equation}

We calculated in \cite{8} the probability of radiation inside a thick target
taking into account the correction term $v(\mbox{\boldmath$\varrho$})$
to the potential $V_c(\mbox{\boldmath$\varrho$})$, see \ref{5}. This
was important for sewing together with Bethe-Heitler cross section
in the region of photon energies where influence of the multiple
scattering is very weak ($\nu_1 \ll 1$). The contribution of boundary
photons was calculated without the correction term
$v(\mbox{\boldmath$\varrho$})$.

In the case when a target has intermediate thickness ($l \sim l_f$)
mentioned separation of contributions becomes senseless. We consider
this case neglecting by the correction term $v(\mbox{\boldmath$\varrho$})$.
The typical mean value we have to calculate (see (\ref{1})-(\ref{5})) is
\begin{eqnarray}
&&\hspace{-8mm}\left<0\left|\exp(iH_0 t_1) \exp(-iH t_2)\right|0 \right>
\rightarrow \left<0\left|\exp(iH_0 t_1) \exp(-iH_c t_2)\right|0
\right>= \nonumber \\
&&\hspace{-8mm}\left<0\left|\exp(iH_0 t_1)\left|\mbox{\boldmath$\varrho$} \right>
\left<\mbox{\boldmath$\varrho$} \right| \exp(-iH_c t_2)\right|0 \right>=
\int_{}^{}d^2\varrho K_0^{\ast}(0, \mbox{\boldmath$\varrho$}, t_1)
K_c(\mbox{\boldmath$\varrho$}, 0, t_2),
\label{9}
\end{eqnarray}
where $\displaystyle{H_0={\bf p}^2,~H=H_0+V(\mbox{\boldmath$\varrho$}),~
H_c=H_0+V_c(\mbox{\boldmath$\varrho$})}$. The Green functions
\newline $K_c(\mbox{\boldmath$\varrho$}_1, \mbox{\boldmath$\varrho$}_2, t)$
and $K_0(\mbox{\boldmath$\varrho$}_1, \mbox{\boldmath$\varrho$}_2, t)$
are defined in \cite{8} (see Eqs.(2.27), (2.24)). Caring out the
calculations (some results obtained in Sections 4, 6 \cite{8}) we find
for the spectral probability of radiation
\begin{eqnarray}
&& \frac{dw}{d\omega}=\frac{\alpha}{\pi \omega}
\sum_{k=1}^{4}\left[-r_1 {\rm Im}~F_k^{(1)} +
r_2 {\rm Re}~F_k^{(2)}\right]; \nonumber \\
&& F_1^{(m)} = F_3^{(m)} =\int_{0}^{\infty} dt_1 \int_{0}^{T} dt_2
{\rm e}^{-it_1}\left[\left(t_1+t_2\right)^{-m}{\rm e}^{-it_2}-
N_1^m{\rm e}^{-i \kappa t_2} \right], \nonumber \\
&& F_2^{(m)} = \int_{0}^{T} dt (T-t)
\left[\left(t-i0\right)^{-m}{\rm e}^{-it}-
N_2^m{\rm e}^{-i \kappa t} \right], \nonumber \\
&& F_4^{(m)} =\int_{0}^{\infty} dt_1 \int_{0}^{\infty} dt_2
{\rm e}^{-i\left(t_1+t_2 \right)}\left[\left(t_1+t_2+T\right)^{-m}
{\rm e}^{-iT}-
N_4^m{\rm e}^{-i \kappa T} \right]
\label{10}
\end{eqnarray}
where
\begin{eqnarray}
&&\hspace{-8mm} N_1=\frac{\nu}{\sinh \nu t_2 + \nu t_1 \cosh \nu t_2},\quad
N_2 = \frac{\nu}{\sinh \nu (t-i0) }, \nonumber \\
&&\hspace{-8mm} N_4=\frac{\nu}{\left(1+\nu^2t_1t_2 \right)\sinh \nu T +
\nu \left(t_1+t_2 \right) \cosh \nu T},\quad
\nu = \frac{1+i}{\sqrt{2}}\nu_0.
\label{11}
\end{eqnarray}
Note that in left-hand side of formula (\ref{10}) $m$ is an index, while
in right-hand side $m$ is a degree of a relevant functions.
The functions $F_k^{(1,2)}~, 1 \leq k \leq 4$ are respectively the
contributions of four domains of integration over $t_1$ and $t_2$
(see \cite{8}, section 4):
four domains:
\begin{enumerate}
\item $t_1 \leq 0,~0 \leq t_2 \leq T$;
\item $ 0 \leq t_1 \leq T,~0 \leq t_2 \leq T$;
\item $0 \leq t_1 \leq T,~ t_2 \geq T$;
\item $t_1 \leq 0,  t_2 \geq T$;
\end{enumerate}
in two more domains $t_{1,2} \leq 0$ and
$t_{1,2} \geq T$ an electron is moving entirely free and there is no
contribution from these domains.

Rearranging the subtraction terms in Eqs.(\ref{10}), (\ref{11})
we present the spectral probability of radiation as
\begin{eqnarray}
&&\hspace{-6mm} \frac{dw}{d\omega}=\frac{\alpha}{\pi \omega}
\sum_{k=1}^{5}\left[-r_1 {\rm Im}~J_k^{(1)} +
r_2 {\rm Re}~J_k^{(2)}\right]; \nonumber \\
&&\hspace{-6mm} J_1^{(m)} = J_3^{(m)} =\int_{0}^{\infty} dt_1 \int_{0}^{T} dt_2
{\rm e}^{-i\left(t_1+\kappa t_2 \right)}\left[\left(t_1+t_2\right)^{-m}
-N_1^m \right], \nonumber \\
&&\hspace{-6mm} J_2^{(m)} = \int_{0}^{T} dt (T-t){\rm e}^{-i \kappa t}
\left[t^{-m}-N_2^m \right], \nonumber \\
&& J_4^{(m)} =\int_{0}^{\infty} dt_1 \int_{0}^{\infty} dt_2
{\rm e}^{-i\left(t_1+t_2+\kappa T \right)}\left[\left(t_1+t_2+T\right)^{-m}
-N_4^m \right] \nonumber \\
&&\hspace{-6mm} J_5^{(m)}(T)=2 \int_{0}^{\infty}dt_1\int_{0}^{T}dt_2
\frac{{\rm e}^{-it_1}}
{\left(t_1+t_2 \right)^m}\left({\rm e}^{-it_2}-{\rm e}^{-i\kappa t_2} \right)
+\int_{0}^{T}dt \frac{(T-t)}{(t-i0)^{m}}
\left({\rm e}^{-it}-{\rm e}^{-i\kappa t} \right) \nonumber \\
&&\hspace{-6mm} + \int_{0}^{\infty}dt_1\int_{0}^{\infty}dt_2
\frac{{\rm e}^{-i\left(t_1+t_2 \right)}}
{\left(t_1+t_2 +T \right)^m}\left({\rm e}^{-iT}-{\rm e}^{-i\kappa T} \right).
\label{12}
\end{eqnarray}
The functions $J_5^{(m)}(T)$ we split into two parts:
\begin{equation}
\displaystyle{J_5^{(m)}(T)=J_5^{(m)}(\infty)+j_5^{(m)}}.
\label{13}\end{equation}
Here $J_5^{(m)}(\infty)$ is the sum of the two first terms in the expression
for $J_5^{(m)}(T)$ where the upper limit of integration $T$ is substituted
$T \rightarrow \infty$. So, the expression for $j_5^{(m)}$ is
\begin{eqnarray}
&&\hspace{-6mm}j_5^{(m)}(T)=2 \int_{0}^{\infty}dt_1\int_{T}^{\infty}
dt_2 \frac{{\rm e}^{-it_1}}
{\left(t_1+t_2 \right)^m}\left({\rm e}^{-it_2}-{\rm e}^{-i\kappa t_2} \right)
+\int_{T}^{\infty}dt \frac{(T-t)}{(t-i0)^{m}}
\left({\rm e}^{-it}-{\rm e}^{-i\kappa t} \right) \nonumber \\
&&\hspace{-6mm} + \int_{0}^{\infty}dt_1\int_{0}^{\infty}dt_2
\frac{{\rm e}^{-i\left(t_1+t_2 \right)}}
{\left(t_1+t_2 +T \right)^m}\left({\rm e}^{-iT}-
{\rm e}^{-i\kappa T} \right).
\label{14}
\end{eqnarray}
In the two-fold integrals in expressions for $J_5^{(m)}(\infty)$ and
$j_5^{(m)}$ we replace the variables to $t=t_1+t_2$ and $t_2$ take integrals
over $t_2$. After this the expression for $J_5^{(1)}(\infty)$ contains
the integral
\begin{equation}
\int_{0}^{\infty}\frac{dt}{t-i0}\left({\rm e}^{-it}-
{\rm e}^{-i\kappa t} \right) = \int_{0}^{\infty} \frac{dt}{t}
\left(\cos t- \cos \kappa t \right)
+ \frac{1}{2}\int_{-\infty}^{\infty}\frac{dt}{t-i0}
\left({\rm e}^{-it}-{\rm e}^{-i\kappa t} \right).
\label{15}
\end{equation}
The first term on the right-hand side is the Froullani integral equal to
$\ln \kappa$. In the second term the integration contour can be closed
in the lower half-plane, where the integrand has no singularities, so that
this integral vanishes. Using the above analysis we have
\begin{eqnarray}
&& -{\rm Im}~J_5^{(1)}(\infty) \equiv J_{tr}^{(1)}=
2 \int_{0}^{\infty}\frac{dt}{t}\left[t \sin t - \frac{1}{\kappa-1}
\left(\cos t- \cos \kappa t \right) \right]- \nonumber \\
&& \int_{0}^{\infty} dt
\left(\sin t- \sin \kappa t \right)= 1+\frac{1}{\kappa}-
\frac{2}{\kappa-1} \ln \kappa
\label{16}
\end{eqnarray}
Similarly we have
\begin{eqnarray}
&& {\rm Re}~J_5^{(2)}(\infty) \equiv J_{tr}^{(2)}=
2 \int_{0}^{\infty}\frac{dt}{t^2}\left[t \cos t - \frac{1}{\kappa-1}
\left(\sin \kappa t- \sin t \right) \right]- \nonumber \\
&& \int_{0}^{\infty} dt
\left(\cos t- \cos \kappa t \right)= \left(1+\frac{2}{\kappa-1} \right)
 \ln \kappa-2,
\label{17}
\end{eqnarray}
where integration by parts is fulfilled.

It is easy to check directly that the sum of the terms in $j_5^{(m)}$
which don't contain the parameter $\kappa$ vanishes. Because of this
we can write $j_5^{(m)}$ in the form
\begin{eqnarray}
&& j_5^{(m)} = \int_{T}^{\infty}\frac{dt}{t^m}
\left[\frac{2i}{\kappa-1} E_{-}(t)+(T-t)E_{+}(t) \right] \nonumber \\
&& E_{\pm}(t) = {\rm e}^{-i\kappa t} \pm
{\rm e}^{-i\left(t+(\kappa-1)T \right)}
\label{18}
\end{eqnarray}
Integrals in (\ref{18}) can be expressed in terms of integral sine
${\rm si}(x)$ and integral cosine ${\rm ci}(x)$ (see \cite{8a}):
\begin{eqnarray}
&& -{\rm Im}~j_5^{(1)} = \frac{2}{\kappa-1}A(\kappa, T) -
T B_1(\kappa, T) -\left(1+\frac{1}{\kappa} \right) \cos \kappa T \nonumber \\
&& {\rm Re}~j_5^{(2)} = - \left(1+\frac{2}{\kappa-1}\right) A(\kappa, T) +
T B_2(\kappa, T) + 2 \cos \kappa T,
\label{19}
\end{eqnarray}
where
\begin{eqnarray}
&& A(\kappa, T)={\rm ci}(\kappa T) - \cos \varphi~{\rm ci}(T)
+\sin \varphi~{\rm si}(T), \nonumber \\
&& B_1(\kappa, T) =
{\rm si}(\kappa T) + \cos \varphi~{\rm si}(T)
+\sin \varphi~{\rm ci}(T), \nonumber \\
&& B_2(\kappa, T) =
\kappa~{\rm si}(\kappa T) + \cos \varphi~{\rm si}(T)
+\sin \varphi~{\rm ci}(T), \quad \varphi=\left(\kappa-1 \right)T
\label{20}
\end{eqnarray}

In the expression for the spectral probability of radiation in form
(\ref{12}) in the limit $\nu_0 T \ll 1$
contribution of terms $J_k^{m}~$(k=1,2,3) becomes small
(it is proportional to powers of $\nu_0 T$). Moreover, for
$\kappa T \geq 1$ the main contribution
gives term $J_5^{m}$ which is $\nu$-independent. This term describes
{\em transition radiation} (see (\ref{13}), (\ref{17}), (\ref{19}),
(\ref{20})). The term $J_{tr}^{(2)}$ is known expression for
probability of transition radiation on one boundary, while
$J_{5}^{(2)}$ as a whole describes the transition radiation on the plate
with two boundaries (in the frame of the classical
electrodynamics) and coincides with corresponding results in the
transition radiation theory (see e.g.\cite{9}). Our complete result
in this case gives the probability of transition radiation in high-energy
quantum electrodynamics.

For numerical calculations it is desirable to improve a convergence of the
integrals in (\ref{12}). For example, in the integral $J_1^{(2)}$ we rotate
the integration contour over $t_1$ at an angle -$\pi /2$ and pass on to
the variable -$it_1$. Then we have
\begin{eqnarray}
&& J_1^{(2)} = i\int_{0}^{\infty} dt_1 \exp (-t_1) \int_{0}^{T} dt_2
\left[\frac{1}{(t_1+it_2)^2}-
G_1^2 \right] \exp(-i\kappa t_2), \nonumber \\
&& G_1=\frac{1}{t_1 \cosh \nu t_2 + (i/\nu) \sinh \nu t_2},\quad
\nu = \frac{1+i}{\sqrt{2}}\nu_0, \quad \frac{i}{\nu}=
\frac{1+i}{\sqrt{2}} \frac{1}{\nu_0}.
\label{21}
\end{eqnarray}
Taking integral over $t_1$ we obtain representation of $J_1^{(2)}$
as a single integral which
is more convenient for numerical calculations
\begin{eqnarray}
&& J_1^{(2)} = i\int_{0}^{T} dt \exp(-i\kappa t)\Bigg\{\exp (it)
{\rm Ei}(-it) - \frac{i}{t} \nonumber \\
&& - \frac{1}{\cosh^2 \nu t} \left[\exp \left(\frac{i \tanh \nu t}{\nu}
\right) {\rm Ei} \left(-\frac{i \tanh \nu t}{\nu}
\right) - \frac{i \nu}{\tanh \nu t} \right] \Bigg\},
\label{21a}
\end{eqnarray}
where ${\rm Ei}(z)$ is the exponential
integral function defined as in \cite{8a}.
In calculations one has to use appropriate branch of the function
${\rm Ei}(z)$ in the complex plane.

In the integral $J_4^{(2)}$ we substitute $t_1 \rightarrow -it_1$ and
$t_2 \rightarrow -it_2$ and then replace the variables $t=t_1+t_2, x=t_2$.
The result is
\begin{eqnarray}
&& J_4^{(2)} = \exp (-i\kappa T)\int_{0}^{\infty} dt\exp (-t)
\int_{0}^{t} dx \left[\frac{1}{(t+iT)^2}- G_4^2 \right], \nonumber \\
&& G_4=\frac{1}{\left(1-\nu^2x(t-x) \right)(i/\nu) \sinh \nu T +
t\cosh \nu T}.
\label{22}
\end{eqnarray}

We consider first the case when LPM effect is weak ($\nu_1 \ll 1$).
Then for thickness $T \ll 1/\nu_1$ the transverse shift of the
projectile due to the multiple scattering in a target as a whole
have no influence on coherent effects defined by the phase
$\phi=\omega l(1-{\bf nv})$ in the factor $\exp(-i\phi)$.
We assume here that condition $\nu_1(\omega_p) \geq 1$
(definition of $\omega_p$ see in (\ref{2})) is fulfilled,
that is in the region where $\nu_1 \ll 1$ one has $\omega \gg \omega_p$
and effects of the polarization of a medium are negligible.
This is true for high energies ($\varepsilon \geq 10~GeV$). Indeed,
for the projectile traversing a target
in the case $\nu_1T \ll 1$ an increment of the phase $\phi$ is small
\begin{equation}
\Delta \phi \sim \omega l \vartheta_s^2 \sim
\omega l \frac{\nu_1^2T}{\gamma^2} \sim \nu_1^2 T^2 \ll 1
\label{22a}
\end{equation}
The angle of multiple scattering $\vartheta_s$ is small also comparing
with an characteristic angle of radiation $1/\gamma~(\gamma^2\vartheta_s^2
=\nu_1^2T \ll 1)$. So, in the case $\nu_1 \ll 1,~ \nu_1T \ll 1$
the radiation originates on separate atoms of a target and an interference
on target boundaries is defined by the value $\omega l (1-v)=T$.
At $T \ll 1$ this interference is weak, while at $T \gg 1$ there is
an exponential damping of the interference
terms due to integration over photon emission angles.
Expanding over $\nu_1$ in (\ref{12}) we obtain ($\kappa=1$):
\begin{eqnarray}
&& {\rm Re}~J^{(2)}(T) = {\rm Re}~\sum_{k=1}^{4}J_k^{(2)}(T) \simeq
\frac{\nu_1^2T}{3}\left[1-3T\int_{1}^{\infty}\frac{(x-1)^2}{x^3}
\sin (xT) dx \right]  \nonumber \\
&& =\frac{\nu_1^2T}{3} \left[1+3T\left(\left(1+\frac{T^2}{2} \right)
{\rm si}(T)-2T {\rm ci}(T) +\frac{3}{2} \sin T -
\frac{T}{2} \cos T \right) \right].
\label{22b}
\end{eqnarray}
For case $T \ll 1$
\begin{equation}
{\rm Re}~J^{(2)}(T) \simeq \frac{\nu_1^2T}{3} \left[1-\frac{3\pi}{2}T
+6T^2\left(\ln \frac{1}{T}+1-C \right) \right],
\label{22c}
\end{equation}
and for case $T \gg 1$
\begin{equation}
{\rm Re}~J^{(2)}(T) \simeq \frac{\nu_1^2T}{3} \left(1+6
\frac{\cos T}{T^2} \right).
\label{22d}
\end{equation}
Thus, in the case $\nu_1 \ll 1,~ \nu_1T \ll 1$ the probability of
radiation is defined by Bethe-Heitler formula both for $T \ll 1$ and
for $T \gg 1$. However, for $T \sim 1$ the interference on the target
boundaries may be essential.

When the parameter $\nu_1T$ is large ($\nu_1 \ll 1,~ \nu_1T \gg 1$)
the radiation is formed inside a target and the interference terms are
damping exponentially. In this case formulae derived in \cite{8} for thick
target are applicable. In this case value of separate terms in the
sum for ${\rm Re}~J^{(2)}(T)$ could strongly oscillate:
\begin{eqnarray}
&& {\rm Re}~J_1^{(2)}(T) = {\rm Re}~J_3^{(2)}(T)) \simeq
{\rm Re}~J_1^{(2)}(\infty) - \int_{0}^{\infty}dt_1 \int_{T}^{\infty} dt_2
\frac{\cos (t_1+t_2)}{(t_1+t_2)^2}  \nonumber \\
&& ={\rm Re}~J_1^{(2)}(\infty) - \Delta(T),\quad
{\rm Re}~J_2^{(2)}(T)
\simeq {\rm Re}~J_2^{(2)}(\infty) +\Delta(T),  \nonumber \\
&& {\rm Re}~J_4^{(2)}(T) \simeq \int_{0}^{\infty}dt_1 \int_{0}^{\infty} dt_2
\frac{\cos (t_1+t_2+T)}{(t_1+t_2+T)^2} = \Delta(T), \nonumber \\
&& \Delta(T)=\int_{T}^{\infty}\frac{(t-T)}{t^2}\cos T dt
\simeq - \int_{T}^{\infty} \frac{\sin t}{t^2} dt \simeq -\frac{\cos T}{T^2}.
\label{22e}
\end{eqnarray}
It is seen from (\ref{22e}) that in the sum for ${\rm Re}~J^{(2)}(T)$
the contribution of terms $\Delta(T)$ is canceled exactly. In the
considered case ($\nu_1 \ll 1$) the value ${\rm Re}~J^{(2)}(\infty)$
gives the formula Bethe-Heitler with corresponding corrections. Remind
that in the limit $\nu_1 \rightarrow 0$ the exact Bethe-Heitler formula
can be obtained only if the terms $\propto 1/L$ are taken into account
\cite{8}. The expression for ${\rm Re}~J^{(2)}$ found in Sect.4
of \cite{8} is
\begin{equation}
{\rm Re}~J^{(2)}(\infty) = \frac{\nu_1^2T}{3} \left(1+\frac{1}{6L_1}
-\frac{16 \nu_1^4}{21}\right) - \frac{2\nu_1^4}{21},
\label{22f}
\end{equation}
where $L_1$ and $\nu_1$ are defined in (\ref{4}) and (\ref{6}).

We consider now the case when the LPM effect is strong ($\nu_0 \gg 1$)
and the parameter $T \ll 1$ while the value which characterize the
thickness of a target $\nu_0 T \sim 1$.
Such situation is possible at $\omega \ll \varepsilon$. So, we can omit
terms with $r_1=\omega^2/\varepsilon^2$ and put $r_2 \simeq 2$.
We expand the expressions for $J_k^{(2)}$ in a power series in
$\displaystyle{\frac{1}{\nu_0}}$ and $T$ including linear terms in
$\displaystyle{\frac{1}{\nu_0}}$ and $T$.
The resulting decompositions are
\begin{eqnarray}
&& J_1^{(2)} = J_3^{(2)} \simeq \ln \frac{\nu T}{\tanh \nu T} +
i\Bigg[\left(\frac{\tanh \nu T}{\nu}-T \right)\left(1-C-i\frac{\pi}{2}
\right)   \nonumber \\
&& +\frac{\tanh \nu T}{\nu} \ln \frac{\nu}{\tanh \nu T}
- T \ln \frac{1}{T} + \kappa \left(\frac{2}{\nu}\int_{0}^{\nu T}
\frac{tdt}{\sinh 2t} - T\right) \Bigg], \nonumber \\
&& J_2^{(2)} \simeq \left(1+i \kappa T \right)\ln \frac{\sinh \nu T}{\nu T}
+ 2i\kappa\left(T-\frac{1}{\nu}\int_{0}^{\nu T}dt t
\coth t \right),~ J_4^{(2)} \simeq \exp (-i\kappa T) \nonumber \\
&& \times \Bigg\{2 \ln \tanh \nu T
+\frac{2i}{\nu} \Bigg[\coth \nu T \ln \frac{\nu}{\coth \nu T}-
\tanh \nu T \ln \frac{\nu}{\tanh \nu T} \nonumber \\
&& + \frac{2}{\sinh 2\nu T}\left(1-C-i\frac{\pi}{2} \right) \Bigg]
-\left(1+2iT \right)\left(\ln T+C+i\frac{\pi}{2} \right)-
1 + iT \Bigg\}.
\label{23}
\end{eqnarray}
The presented here expressions for functions $J_1^{(2)},~J_2^{(2)},~J_3^{(2)}$
are not valid in the case $\kappa T \geq 1$. However, in this case
(under condition that $\nu_0 T \ll 1$) the contribution of terms
$J_1^{(2)},~J_2^{(2)},~J_3^{(2)}$ is negligible both in the asymptotic
expressions (\ref{23}) and exact formulae (\ref{12}). This means that
expressions (\ref{23}) may be used at any value $\kappa T$ when
$\nu_0 \gg 1, T \ll 1$.
Substituting obtained asymptotic decompositions into Eq.(\ref{12})
we find for $\kappa T \ll 1$
\begin{eqnarray}
&& \frac{dw}{d\omega}=\frac{2\alpha}{\pi \omega}
\left({\rm Re}~J^{(2)}+J_5^{(2)}\right),~J_5^{(2)}
\simeq \frac{(\kappa-1)^2}{2}\left(\ln \frac{1}{T}+
\frac{1}{2}-C \right), \nonumber \\
&&{\rm Re}~J^{(2)} = {\rm Re}~\sum_{k=1}^{4} J_k^{(2)} \simeq
{\rm Re}~\Bigg\{ \ln (\nu \sinh \nu T)-1-C-
\kappa \frac{\pi T}{4} + \nonumber \\
&& \frac{2i}{\nu \tanh \nu T}\left[\ln (\nu \tanh \nu T)+1-C-
\frac{i\pi}{2} \right]+i\kappa T\left(\ln \frac{\cosh \nu T}{\tanh \nu T}-
\frac{\nu T}{\tanh \nu T} \right) \nonumber \\
&& + \frac{i\kappa}{\nu}\int_{0}^{\nu T} dt \left(\frac{4t}{\sinh 2t}-
\frac{t^2}{\sinh^2 t} \right)\Bigg\}, \quad
\nu = \exp\left(i\frac{\pi}{4}\right)\nu_0.
\label{24}
\end{eqnarray}

For a relatively thick target ($\nu_0T \gg 1$) we have from (\ref{24})
\begin{eqnarray}
&& {\rm Re}~J^{(2)} \simeq \ln \nu_0 -1 -C -
\ln 2+\frac{\sqrt{2}}{\nu_0}\left(\kappa\frac{\pi^2}{24}+
\ln \nu_0 +1-C+\frac{\pi}{4} \right) \nonumber \\
&& + \frac{\nu_0T}{\sqrt{2}}\left(1-\frac{\pi \kappa}{2\sqrt{2}\nu_0} \right)
\label{25}
\end{eqnarray}
Here the terms without $T$ are the contribution of boundary photons
(formula (4.14) of \cite{8}) while the term $\propto T$ gives in
(\ref{25}) the probability of radiation inside target (with correction
$\sim \kappa/\nu_0$ but without corrections $\sim 1/L$). The relative
value of the last corrections at $\nu_0 \gg 1$ is (Eq.(2.45) of \cite {8})
\begin{equation}
r=\frac{1}{2L(\varrho_c)}\left(\ln 2-C+\frac{\pi}{4} \right) \simeq
\frac{0.451}{L(\varrho_c)}.
\label{25a}
\end{equation}

In the limiting case when a target is very thin and $\nu_0 T \ll 1$ but when
$\nu_0^2 T \gg 1$ we have from (\ref{24})
\begin{eqnarray}
&& {\rm Re}~J^{(2)} \simeq \left(1+\frac{2}{\nu_0^2T} \right)
\left[\ln (\nu_0^2T) +1 -C \right] - 2 +\delta, \nonumber \\
&& \delta= \frac{(\nu_0 T)^4}{180}+\frac{2(\nu_0 T)^2}{45}T\left(
\ln \nu_0^2 T - C\right)-\frac{\kappa T}{6}(\nu_0 T)^2.
\label{26}
\end{eqnarray}
The terms without $\delta$ in this expression coincide
with formula (5.15) of \cite{8} (up to terms
$\propto~C/L_t$).

In the photon energy region where $\nu_0T \ll 1$ the contribution of
the terms $J_k^{(m)}$~(k=1,2,3) is very small
($\sim \delta$) and decreases
with photon energy reduction ($\propto \omega$), so that in the
spectral distribution of radiation only the terms $J_4^{(m)}, J_5^{(m)}$
contribute. We consider now the function $J_4^{(2)}$ in the case when
$(1+\nu_0)T \ll 1$ and the parameter $\nu_0^2T$, which characterizes
the mean square angle of the multiple scattering in a target as a whole,
has an arbitrary value. Under the mentioned conditions the function $N_4$
in (\ref{11}) may be written as
\begin{equation}
N_4^2 \simeq \left(\nu^2 T t_1 t_2+t_1+t_2 \right)^{-2}=
-\int_{0}^{\infty}dx x \exp \left[-ix\left(\nu^2 T t_1 t_2+t_1+
t_2 \right) \right].
\label{27}
\end{equation}
Substituting this expression in (\ref{12}) we find
\begin{equation}
J_4^{(2)}{\rm e}^{i\kappa T} \simeq \int_{0}^{\infty}dx x \int_{0}^{\infty}dt_1
\int_{0}^{\infty}dt_2 \exp \left(-i(1+x)(t_1+t_2) \right)
\left[\exp \left(-ix\nu^2 T t_1 t_2 \right) -1\right].
\label{28}
\end{equation}
Making the substitution of variables
\[
t_{1,2} \rightarrow -it_{1,2},\quad x \rightarrow \frac{x}{t_1t_2}
\]
we obtain
\begin{eqnarray}
&& \hspace{-4mm} J_4^{(2)}{\rm e}^{i\kappa T} = \int_{0}^{\infty}dx x
\int_{0}^{\infty}\frac{dt_1}{t_1^2}
\int_{0}^{\infty}\frac{dt_2}{t_2^2} \exp \left(-\left(1+
\frac{x}{t_1t_2}\right)(t_1+t_2) \right)
\left[1-\exp \left(-x\nu_0^2 T \right) \right] \nonumber \\
&& \hspace{-4mm} = 4 \int_{0}^{\infty}dx K_1^2\left(2\sqrt{x} \right)
\left[1-\exp \left(-x\nu_0^2 T \right) \right] =2\int_{0}^{\infty}d\varrho
\varrho K_1^2(\varrho)\left[1-\exp \left(-k\varrho^2 \right)
\right], \nonumber \\
&& \hspace{-4mm} 4k=\nu_0^2T,
\label{29}
\end{eqnarray}
where $K_1(\varrho)$ is the modified Bessel function.
Formula (\ref{29}) corresponds at $\kappa=1$ to result for a thin
target obtained in \cite{8} (see Eq.(5.7)) without terms $\propto 1/L$.
Since the dependence on the parameter $\kappa$ is contained in (\ref{29})
as a common phase multiplier $\exp (-i\kappa T)$, one can write more
accurate expression for $J_4^{(2)}$ (with terms $\propto 1/L$) using
the results of \cite{8}(see Eq.(5.9)):
\begin{eqnarray}
&& J_4^{(2)}=2{\rm e}^{-i\kappa T} \int_{0}^{\infty}d\varrho
\varrho K_1^2(\varrho)\left[1-\exp \left(-V(\varrho)T \right)
\right], \nonumber \\
&& V(\varrho)T=\frac{\pi Z^2 \alpha^2 n l \varrho^2}{m^2}
\left(\ln \frac{4a_{s2}^2}{\lambda^2\varrho^2} - 2 C\right).
\label{30}
\end{eqnarray}
For the case $\nu_0^2T \gg 1$ it has the form
\begin{eqnarray}
&& {\rm e}^{i\kappa T}~J_4^{(2)} = \left(1+\frac{1}{2k} \right)
\left[\ln 4k +1 -C \right] - 2 +\frac{C}{L_t},\quad 4k=\nu_0^2T, \nonumber \\
&& k= \frac{\pi Z^2 \alpha^2 n l}{m^2}L_t, \quad
L_t= \ln \frac{4a_{s2}^2}{\lambda_c^2 \varrho_t^2}-2C,\quad
k(\varrho_t)\varrho_t^2=1
\label{31}
\end{eqnarray}
In the case when parameter $k$ is not very high one has to use
an exact expression found in \cite{8} (formula (5.7)).
For $k \ll 1$
one can expand the exponent in the integrand of (\ref{30}). Then we find
\begin{equation}
{\rm e}^{i\kappa T}J_4^{(2)} = \frac{\nu_1^2T}{3}
\left(1+\frac{1}{6L_1} \right),\quad \nu_1^2T=
\frac{4\pi Z^2 \alpha^2nl}{m^2}L_1.
\label{32}
\end{equation}
At $\kappa T \ll 1$ the spectral distribution of probability is
\begin{equation}
\frac{dw}{d\omega}=\frac{2\alpha}{3\pi \omega}\nu_1^2 T\left(1+
\frac{1}{6L_1} \right)\left(1-\frac{\omega}{\varepsilon}\right).
\label{33}
\end{equation}
This is the Bethe-Heitler formula for not very hard photons (terms
$\displaystyle{\propto
\left(\frac{\omega}{\varepsilon}\right)^2}$ are omitted).

When a photon energy decreases, the parameter $\kappa$ increases as well as
the combination $\kappa T \propto 1/\omega$, while the value
$(\nu_0 T)^2$ decreases $\propto \omega$. Just this value defines an
accuracy of Eq.(\ref{30}). Using Eqs.(\ref{13})-(\ref{20}) at $T \ll 1,
\kappa T \geq 1$ we find for the probability of the transition radiation
following expression
\begin{eqnarray}
&& \frac{dw_{tr}}{d\omega} \simeq \frac{2\alpha}{\pi}
\Bigg\{\left(1+\frac{2}{\kappa-1} \right)
\Bigg[\ln \kappa -{\rm ci}(\kappa T) + \cos (\kappa T)(\ln T+C) \nonumber \\
&& + \frac{\pi}{2} \sin (\kappa T) \Bigg] + \kappa T {\rm si}(\kappa T)
-4 \sin^2 \frac{\kappa T}{2} \Bigg\}.
\label{34}
\end{eqnarray}
In the limiting case $\kappa T \gg 1$ the probability (\ref{34}) turns
into standard probability of the transition radiation with oscillating
additions
\begin{equation}
\frac{dw_{tr}}{d\omega}=\frac{2\alpha}{\pi}\left[J_{tr}+\cos (\kappa T)
\left(\ln T +C+1 \right)+\frac{\pi}{2}\sin (\kappa T) \right]
\label{34a}
\end{equation}
Note, that there is a qualitative difference in a behaviors of
interference terms in Eqs.(\ref{22d}) and (\ref{34a}). In the former an
amplitude of oscillation with $\omega$ increase decreases as $1/\omega$
whilst in the latter the corresponding amplitude weakly (logarithmically)
increases with $\omega$ decrease.

From the above analysis follows that in the case when
$\nu_0 T \ll 1$ ($\nu_0 \gg 1$)
the spectral distribution of probability of
radiation with the polarization of a medium
taken into account has the form
\begin{equation}
\frac{dw}{d\omega} = \frac{dw_{tr}}{d\omega} +\cos (\kappa T)
\frac{dw_{th}}{d\omega},
\label{35}
\end{equation}
where $dw_{th}/d\omega$ is the spectral distribution of probability
of radiation in a thin target without regard for the polarization of
a medium. In the case $4k = \nu_0^2 T \gg 1$ the probability
$dw_{th}/d\omega$ is defined by Eq.(\ref{31}) and for the case
$k \ll 1$ it is defined by Eq.(\ref{32}). More accurate representation
of the probability of radiation $dw_{th}/d\omega$
may be obtained using Eq.(\ref{30}). It follows from Eqs.(\ref{34a}) and
(\ref{35}) that if we make allowance for multiple scattering
at $\kappa T \gg 1$ this results in decreasing of oscillations of
the transition radiation probability by magnitude of the bremsstrahlung
probability in a thin target.

\section{A qualitative behavior of the spectral intensity of radiation}
\setcounter{equation}{0}

We consider the spectral intensity of radiation for the energy of
the initial electrons when the LPM suppression of the intensity of radiation
takes place for relatively soft energies of photons:
$\omega \leq \omega_c \ll \varepsilon$:
\begin{equation}
\displaystyle{\nu_1(\omega_c)=1,\quad \omega_c=\frac{16\pi Z^2 \alpha^2}{m^2}
\gamma^2 n \ln \frac{a_{s2}}{\lambda_c}},
\label{36}
\end{equation}
see Eqs.(\ref{4}), (\ref{5}), (\ref{6}) and (\ref{8}).
This situation corresponds to
the \newline experimental conditions.

A ratio of a thickness of a target and the formation length of
radiation (\ref{7}) is an important characteristics of the process.
This ratio may be written as
\begin{eqnarray}
&& \beta(\omega)=T\left(\nu_0+\kappa \right)\simeq T_c\left[
\frac{\omega}{\omega_c}+\sqrt{\frac{\omega}{\omega_c}}+
\frac{\omega_p^2}{\omega \omega_c} \right], \nonumber \\
&&T=\frac{l \omega}{2\gamma^2},\quad
\omega_p=\omega_0 \gamma,\quad
T_c \equiv T(\omega_c)
\simeq \frac{2\pi}{\alpha}\frac{l}{L_{rad}},
\label{37}
\end{eqnarray}
where we put that $\displaystyle{\nu_0
\simeq \sqrt{\frac{\omega_c}{\omega}}}$.
Below we assume that $\omega_c \gg \omega_p$ which is true under
the experimental conditions.

If $\beta(\omega_c)=2T_c \ll 1$
then at $\omega=\omega_c$ a target is thin and the Bethe-Heitler
spectrum of radiation is valid at $\omega \leq
\omega_c$ in accordance with Eq.(\ref{30}) since
$4k=\nu_0^2T=T_c \ll 1$. This behavior of the
spectral curve will continue with $\omega$
decrease until photon energies where a contribution of the
transition radiation become essential. In this case the
spectral distribution of radiation has the form (\ref{35}) for all $\omega$
\begin{equation}
\frac{dw}{d\omega} = \frac{dw_{tr}}{d\omega} +\cos (\kappa T)
\frac{dw_{BH}}{d\omega},
\label{38}
\end{equation}
Since for soft photons ($\omega \ll \varepsilon$)
\begin{equation}
\frac{dI}{d\omega} = \frac{2\alpha}{\pi}\left[J_5^{(2)} +
\frac{T_c}{3}\left(1+\frac{1}{6L_1} \right)\cos (\kappa T) \right]
\label{39}
\end{equation}
and $T_c/3 \ll 1$ a contribution of the transition radiation become visible
already at $\kappa T \ll 1$. For $\omega > \omega_c~(T_c \ll 1)$ the
probability of radiation is defined by (\ref{22b})-(\ref{22d}). In this case
a considerable distinction from Bethe-Heitler formula will be in
the region $\omega \sim \omega_c/T_c$.

If $\beta(\omega_c) \gg 1~(T_c \gg 1)$ then at $\omega \geq \omega_c$
a target is thick and one has the LPM suppression for $\omega \leq \omega_c$.
There are two opportunities depending on the minimal value of the parameter
$\beta$.
\begin{equation}
\beta_m \simeq \frac{3}{2}T_c
\sqrt{\frac{\omega_1}{\omega_c}},\quad
\omega_1=\omega_p\left(\frac{4\omega_p}{\omega_c}\right)^{1/3},\quad
\beta_m \simeq 2T_c\left(\frac{\omega_p}{\omega_c} \right)^{2/3}.
\label{40}
\end{equation}
If $\beta_m \ll 1$ then for photon energies $\omega > \omega_1$ it will
be $\omega_2$ such that
\begin{equation}
\beta(\omega_2)=1,\quad \omega_2 \simeq \frac{\omega_c}{T_c^2}
\label{41}
\end{equation}
and for $\omega < \omega_2$ the thickness of a target becomes smaller
than the formation length of radiation so that for $\omega \ll \omega_2$
the spectral distribution of the radiation
intensity is described by formula (\ref{30}). In this case for
$4k=\nu_0^2T=T_c \gg 1$  one has (2.31). Under conditions $\kappa T \ll 1,
\omega < \omega_2$ the intensity of radiation is independent of
photon energy $\omega$. It should be noted that due to smallness of the
coefficients in expression for $\delta$ (\ref{26}), such behaviors of the
spectral curve begins at $\omega < \omega_{th} =4\omega_2 \simeq
4\omega_c/T_c^2$. Such behavior of the spectral curve
will continue until photon energies
where one has to take into account the polarization of a medium
and connected with it a contribution of the transition radiation.

At $\beta_m \gg 1$ a target remains thick for all photon energies
and radiation is described in details by formulae of Sections 2 and 3
of \cite{8} where comparison with experimental data was carried out as well.

For very high energies when the LPM effect becomes significant
at $\omega \sim \varepsilon$ Eq.(\ref{36}) should be substituted
by
\begin{equation}
\displaystyle{\omega_c=\frac{16\pi Z^2 \alpha^2}{m^2}
\gamma^2 \left(1-\frac{\omega}{\varepsilon} \right)n 
\ln \frac{a_{s2}}{\lambda_c}},
\label{41a}
\end{equation}
so that
\begin{equation}
\displaystyle{\frac{\omega_c/\varepsilon}{(1-\omega/\varepsilon)}=
\frac{4\pi}{\alpha}\gamma \frac{\lambda_c}{L_{rad}}=r}
\label{41b}
\end{equation}
It is evident that for $\omega < \omega_c$ the radiation losses diminish
(for very rough estimation one can use as reduction factor $r/(1+r)$)
and due to this the radiation length enlarges. Of course, for the electron
energy $\varepsilon=25~GeV$ this effect is very weak (order of $1\%$).
However, for very high energy it becomes quite sizable. For example,
for the electron energy $\varepsilon=500~GeV$ this effect is of the 
order of $16\%$.

There is, in principle, an opportunity to measure the electron
energy (in region of high energies) using the LPM effect. For this one can
measure the spectral curve on a target with thickness a few percent of
$L_{rad}$ and compare the result with the theory prediction \cite{8}.

Existence of the plateau of the spectral curve in a region of
photon energies where a target is thin was found in
\cite{10} within Migdal approach (quantum theory). Recently
this item was discussed in \cite{11} (in classical theory),
\cite{12} and \cite{13}.

\section{Discussion and conclusions}
\setcounter{equation}{0}

In \cite{8} the qualitative analysis of the data \cite{5}-\cite{7}
was performed. It was noted that for targets with thickness $l \geq
2\% L_{rad}$ the formation length of radiation $l_f \ll l$
for any photon energy $\omega$. So, these targets can be considered as
thick targets. The gold targets with thickness $l=0.7\% L_{rad}$
and $l=0.1\% L_{rad}$ are an exception. We calculated energy losses spectra
in these targets for the initial electron energy $\varepsilon=25~GeV$
and $\varepsilon=8~GeV$. The characteristic parameters of radiation for
these cases are given in Table.

In Fig.1(a) results of calculations are given for target with a thickness
\newline $l=0.7\%L_{rad}$ at $\varepsilon=25~GeV$. The curves 1,2,3,4
present correspondingly the functions $J_1^{(2)}, J_2^{(2)},
J_3^{(2)}, J_4^{(2)}$ (\ref{12}).
At $\omega=500~MeV$ the value \newline $\displaystyle{\nu_1 T =
\sqrt{\frac{\omega}{\omega_c}}T_c = 8.4 \gg 1}$, the interference terms
are exponentially small and one can use formulae for a thick target.
In this case the parameter $\nu_1=0.69$ and contribution of boundary photons
($J_b=J_1^{(2)}+J_3^{(2)}+J_4^{(2)}$) is small ($\displaystyle{J_b \simeq
- \frac{2\nu_1^4}{21}}$, see \cite{8}, Eq.(4.16)) and distinction
Bethe-Heitler formula ($J_{BH}=T_c/3=1.94$) from
${\rm Re}~J^{(2)}={\rm Re}~(J_1^{(2)}+J_2^{(2)}+
J_3^{(2)}+J_4^{(2)})$ is of the order
10\% according with asymptotic damping factor
$\displaystyle{\left(1-\frac{16\nu_1^4}{21}\right)}$.

At $\omega < \omega_{th} \simeq 30~MeV$ for the case $T_c \gg 1$
and $\beta_m < 1$ the spectral curve turns into plateau according
with discussion in previous Section. In this photon energy region
the parameter $\nu_0 > 3$ and Eq.(\ref{24}) for a target with
intermediate thickness describes the spectral
probability of radiation with a good accuracy. With the further
photon energy decrease one can use limiting formula (\ref{26}) where
$\nu_0^2 T \simeq 7.4$. Note, that in formulae for
$J_1^{(2)} \div J_4^{(2)}$ the potential $V_c(\mbox{\boldmath$\varrho$})$
(\ref{5}) is used which doesn't include corrections
$\sim 1/L~(v(\mbox{\boldmath$\varrho$}))$. These corrections were calculated
in \cite{8} both thin and thick targets. In our case ($\nu_0 \gg 1,~
\nu_0 T \gg 1$) the expressions with corrections $\sim 1/L$ are given in
(\ref{25a}) and (\ref{31}) for a thick target and a thin target respectively.
Taking into account behaviors of correction in the region $\nu_0 \leq 1$
(curve 2 in Fig.2(a) of \cite{8}) we construct an interpolation factor
for term $\sim 1/L$ with accuracy of order 1\%. The summary curve ($T$) in
Fig.1(a) contains this factor.

The transition radiation contributes in the region $\omega \leq \omega_p$
(function ${\rm Re}~J_5^{(2)}$, curve 5 in Fig.1(a)). When $\kappa T \ll 1$
this curve is described by asymptotic of ${\rm Re}~J_5^{(2)}$ (\ref{24}).
The contribution of the transition radiation increases with $\omega$ decrease
and for $\kappa T \sim 1$ it describes by Eq.(\ref{34}). The contribution
of the multiple scattering diminishes due to interference factor
$\cos (\kappa T)$ in (\ref{35}) (at
\newline $\omega=0.2~MeV,~\kappa T=1.9$).
The curve $T$ in Fig.1(a) gives the summary contribution of the
multiple scattering (the curve $S$, where factor $(1-\omega/\varepsilon)$
is included) and the transition radiation (the curve 5).

In Fig.1(b) results of calculations are given for target with a thickness
\newline $l=0.7\%L_{rad}$ at $\varepsilon=8~GeV$.
The notations are the same as in Fig.1(a).
In this case the characteristic photon energy $\omega_c$ is one order of
magnitude lower than for
$\varepsilon=25~GeV$, so that at $\omega=500~MeV$
the parameter $\nu_1$ is small ($\nu_1^2 \simeq 1/20$). Because of this
the right part of the curve $S$ coincides with a good accuracy with
Bethe-Heitler formula (the Coulomb corrections are included). Note, that for
this electron energy the effect of recoil (factor $(1-\omega/\varepsilon)$)
is more essential. Strictly speaking, a target with a thickness
$0.7\%L_{rad}$ at $\varepsilon=8~GeV$ is not thin target for any
photon energy ($\beta_m=1.6$). However, for bremsstrahlung this target
can be considered as a thin one for $\omega < \omega_{th}=3~MeV$.
Since the polarization of a medium becomes essential in the same region
($\omega_p=1.25~MeV$), the interference factor $\cos (\kappa T)$ in
(\ref{35}) causes an inflection of the
spectral curve $S$ at $\omega \sim 1~MeV$.
The transition radiation grows from the same photon energy $\omega$ and
because of this the total spectral curve $T$ has a minimum at
$\omega \simeq 1~MeV$. As far as there is some interval of energies
between $\omega_p$ and $\omega_{th}$ ($\omega_{th}-\omega_p \sim 3~MeV$),
this minimum is enough wide. Moreover, the value of its
ordinate coincide with a good accuracy with ordinate of the plateau of
the spectral curve $S$ in Fig.1(a) because bremsstrahlung on a thin target
is independent of electron energy (\ref{31}).

In Fig.2(a) results of calculations are given for target with a thickness
\newline $0.1\%L_{rad}$ at $\varepsilon=25~GeV$. The notations are the same as in Fig.1(a).
For this thickness $T_c=0.96$ and one has a thin target starting from
$\omega \leq \omega_c$. So, we have here very wide plateau.
The left edge of the plateau is defined by the
contribution of transition radiation ($\omega \sim \omega_p$). Since
in this case $4k=\nu_1^2 T=T_c \simeq 1$ (see (\ref{30})$\div$(\ref{33})),
one has to calculate the ordinate of the plateau using the exact formula
for a thin target (\ref{30}). The same ordinate has the plateau for
electron energy $\varepsilon=8~GeV$ (Fig.2(b)). However, a width of
the plateau for this electron energy is more narrow ($1~MeV \div 20~MeV$)
due to diminishing of the interval between $\omega_p$ and $\omega_c$.
For $\omega > \omega_c$ the formation length of radiation becomes shorter
than target thickness ($T=T_c\omega/\omega_c > 1$) and the parameter
$\nu_1$ decreases. The value $\nu_1 T=T_c\sqrt{\omega/\omega_c}$ increases
with $\omega$ growth. A target becomes thick
and the spectral curves is described by the Bethe-Heitler formula
in Fig.2(b) starting from photon energy $\omega \sim 100~MeV$.
In Fig.2(b) the contributions of separate terms into ${\rm Re}~J^{(2)}$
are shown as well.
Their behavior at $\omega > \omega_c$ is described quite satisfactory
by formulae (\ref{22e})-(\ref{22f}) (see also discussion at their derivation).

We compared our calculations with experimental data \cite{7}. The curves $T$
in Fig.1,2 give theory prediction (no fitting parameters !)
in units $2\alpha/\pi$. We recalculated data according with given in
\cite{7} procedure
\begin{equation}
\left(\omega\frac{dw}{d\omega} \right)_{exp}=\frac{l}{L_{rad}}
\frac{N_{exp}}{k},\quad k=0.09
\label{42}
\end{equation}
It is seen that in Fig.1(a) there is a perfect agreement of the theory and
data. In Fig.1(b) there a overall difference: data
is order of $10\%$ are higher
than theory curve. For photon energy $\omega=500~MeV$ the theory coincide
with Bethe-Heitler formula (with the Coulomb corrections)
applicable for this energy. Note that just for this case it was
similar problem with normalization of data matching with
the Migdal Monte Carlo simulation ($+12.2\%$, see Table II in \cite{7}).

For thickness $l=0.1\%~L_{rad}$ there is a qualitative difference between
our theory prediction and Monte Carlo simulation in \cite{7}. There was a
number of experimental uncertainties associated with this target.
Nevertheless, we show data for $\varepsilon=25~GeV$ which are
lying higher than theory curve.

For target with a thickness $l=0.7\%L_{rad}$ at $\varepsilon=25~GeV$ data was
compared with calculation in \cite{13} (the Coulomb corrections 
were discarded). After arbitrary diminishing
of calculated value by $7\%$ it was found excellent agreement. It is
seen from the above analysis that this subtraction can be considered as
taking account of the Coulomb corrections contribution.

We would like to thank S. Klein for useful comments about data.

\newpage

{\bf Figure captions}

\vspace{15mm}
\begin{itemize}

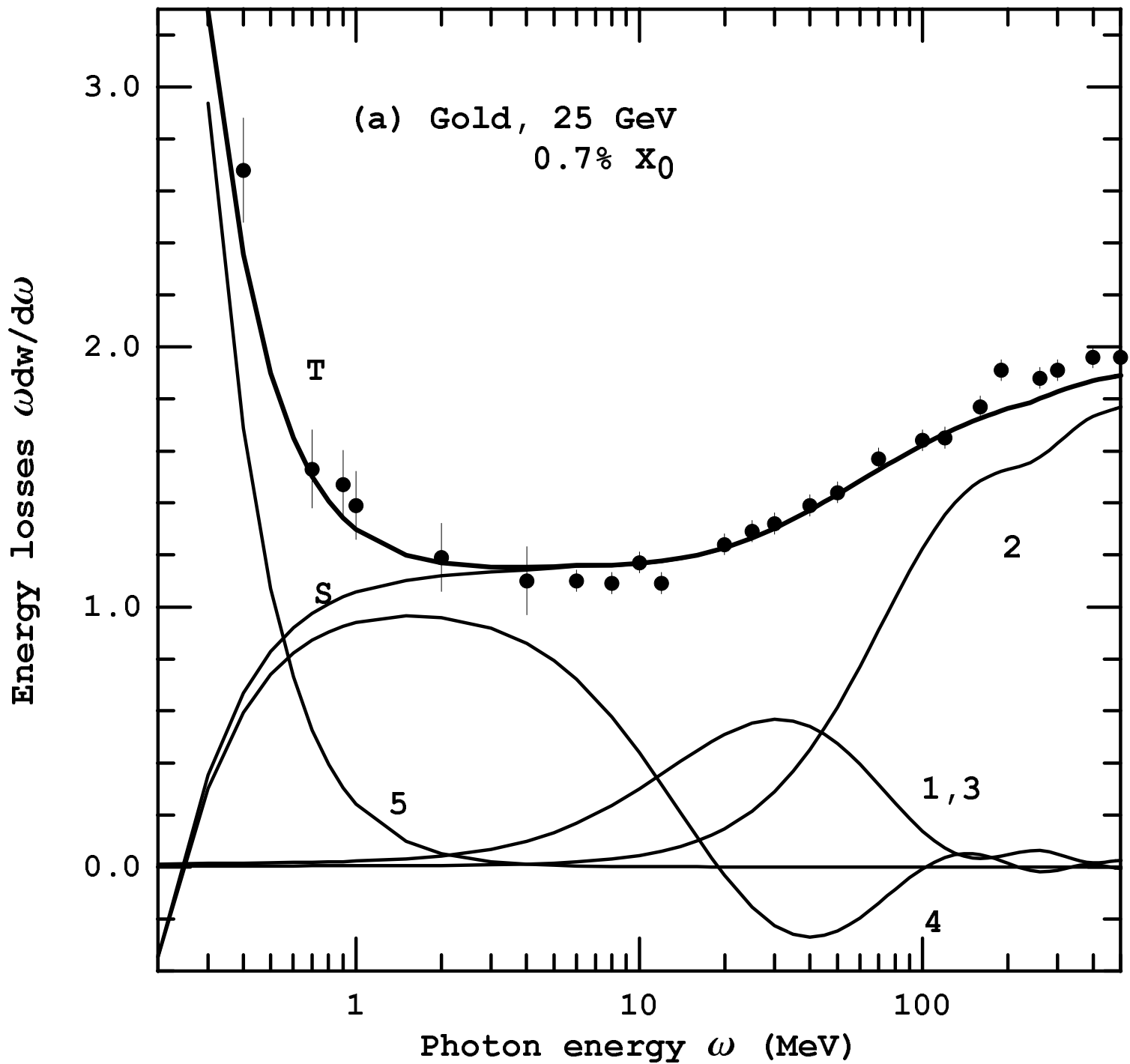
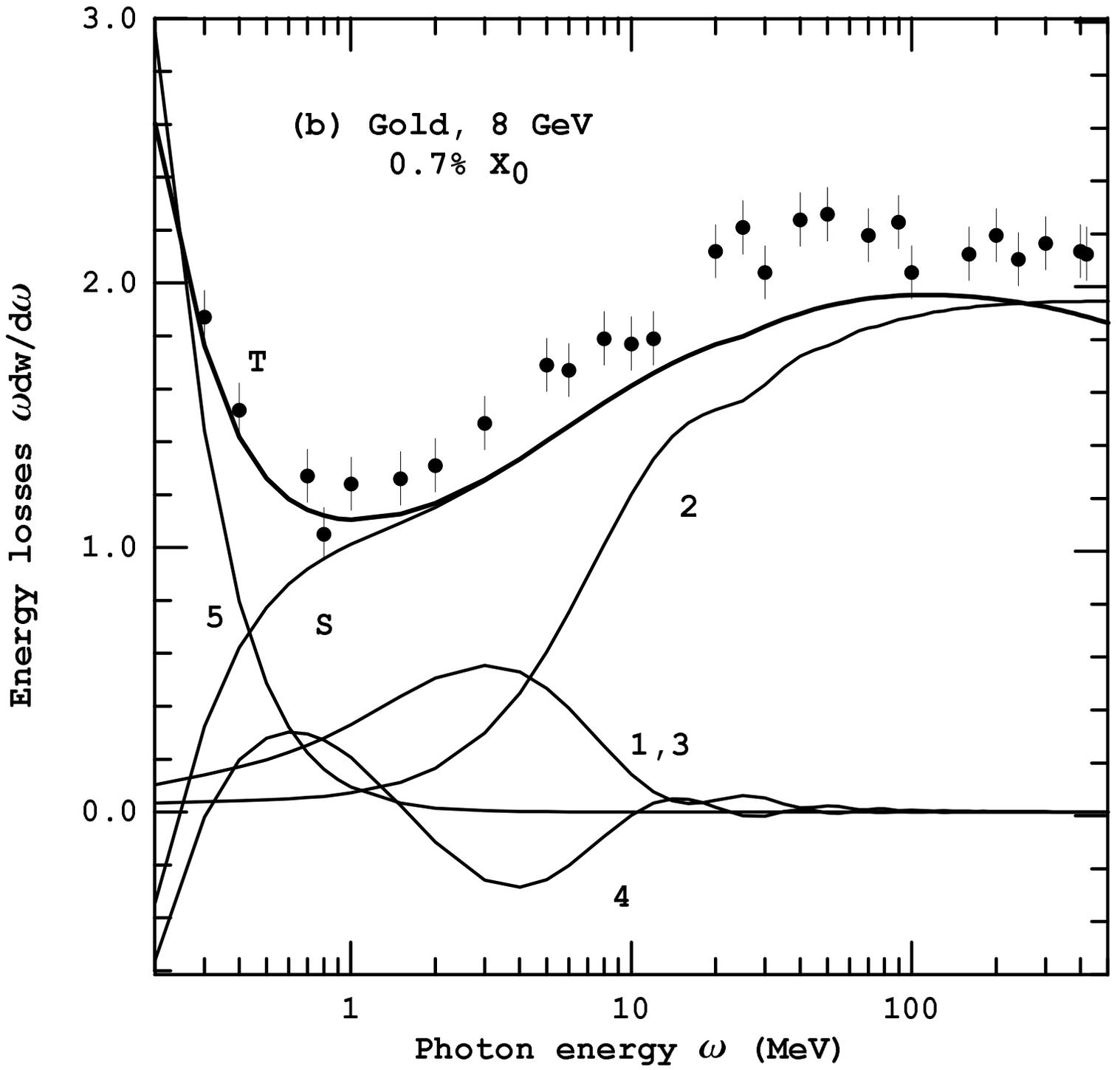
\item {\bf Fig.1} The energy losses
$\displaystyle{\omega \frac{dW}{d\omega}}$
in gold with thickness $l=0.023~mm$ in units
$\displaystyle{\frac{2\alpha}{\pi}}$,
((a) is for the initial electrons energy $\varepsilon=25~GeV$ and (b)
is for $\varepsilon=8~GeV$).
The Coulomb corrections and the polarization of a medium are included.
\begin{itemize}
\item Curve 1 is the contribution of the term ${\rm Re}~J_1^{(2)}=
{\rm Re}~J_3^{(2)}$;
\item curve 2 is the contribution of the term ${\rm Re}~J_2^{(2)}$;
\item curve 4 is the contribution of the term ${\rm Re}~J_4^{(2)}$, all
(\ref{12});
\item curve S is the sum of the previous contributions ${\rm Re}~J^{(2)}$;
\item curve 5 is the contribution of the boundary photons (\ref{19});
\item curve T is the total prediction for the radiation energy losses .
\end{itemize}
Experimental data from Fig.12 of \cite{7}.

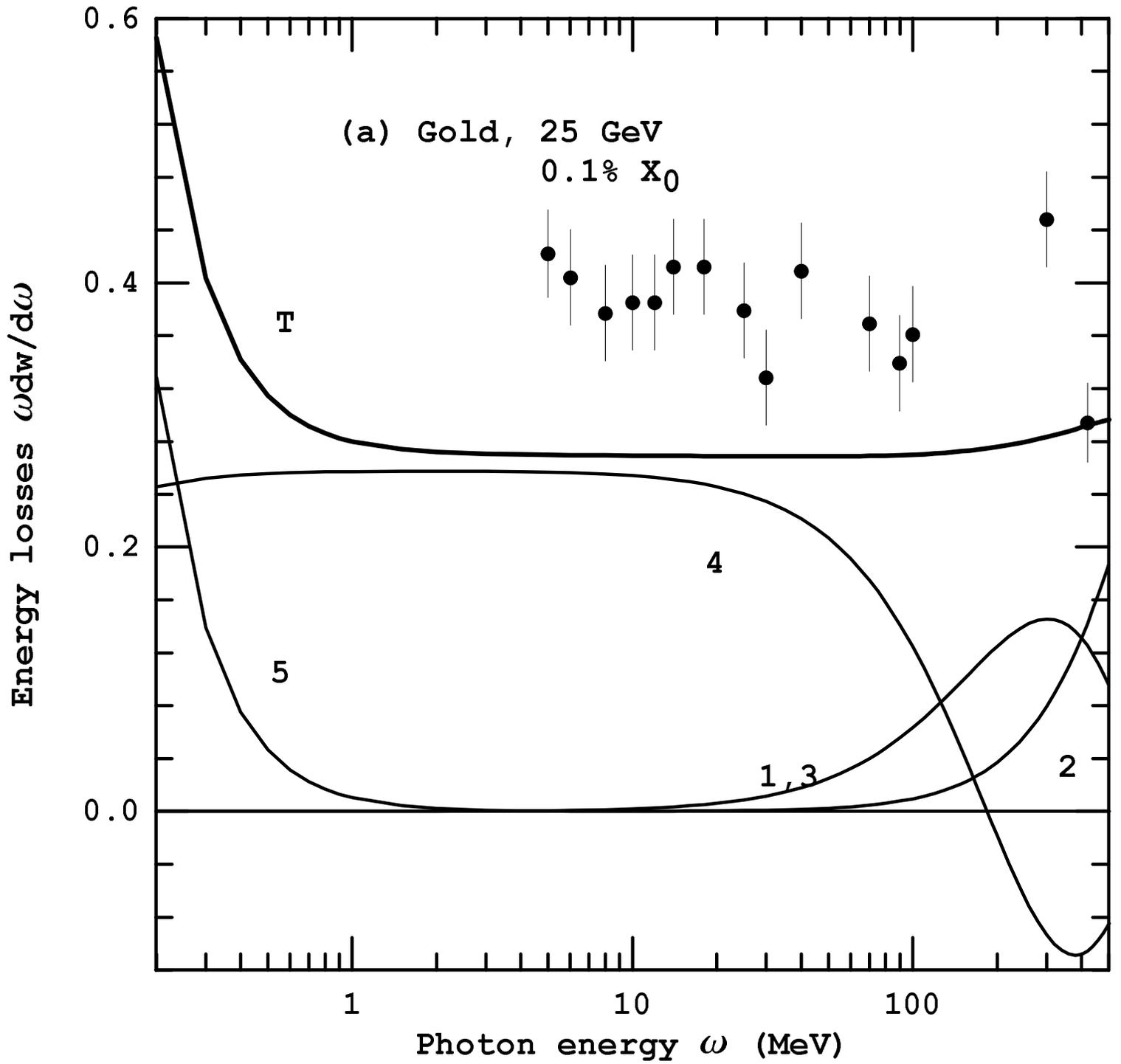
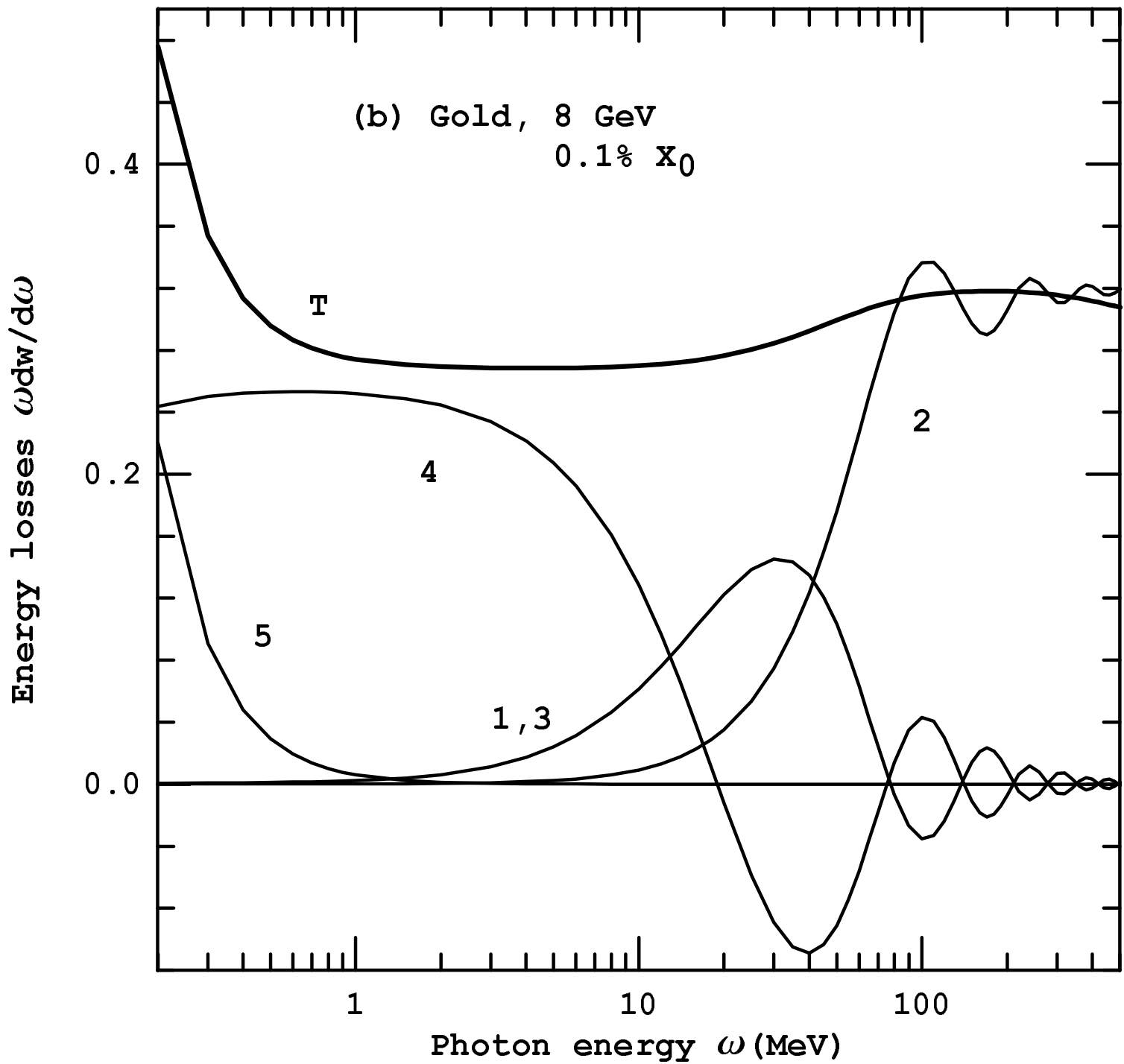
\item {\bf Fig.2} The energy losses
$\displaystyle{\omega \frac{dW}{d\omega}}$
in gold with thickness $l=0.0038~mm$ in units
$\displaystyle{\frac{2\alpha}{\pi}}$,
((a) is for the initial electrons energy $\varepsilon=25~GeV$ and (b)
is for $\varepsilon=8~GeV$).
The Coulomb corrections and the polarization of a medium are included.
\begin{itemize}
\item Curve 1 is the contribution of the term ${\rm Re}~J_1^{(2)}=
{\rm Re}~J_3^{(2)}$;
\item curve 2 is the contribution of the term ${\rm Re}~J_2^{(2)}$;
\item curve 4 is the contribution of the term ${\rm Re}~J_4^{(2)}$, all
(\ref{12});
\item curve S is the sum of the previous contributions ${\rm Re}~J^{(2)}$;
\item curve 5 is the contribution of the boundary photons (\ref{19});
\item curve T is the total prediction for the radiation energy losses .
\end{itemize}
Experimental data from Fig.13 of \cite{7}.

\end{itemize}

\newpage

\begin{table}
\begin{center}
{\sc TABLE}~
{Characteristic parameters of the radiation process}\\
{in gold with the thickness $l = 0.7\% L_{rad}$ and $l = 0.1\% L_{rad}$}
{\newline (all photon energies $\omega$ are in MeV)}
\end{center}
\begin{center}
\begin{tabular}{*{9}{|c}|}
\hline
$\varepsilon~(GeV)$ & $\omega_c~$ & $\omega_p~$& $T_c(0.7)$&
$T_c(0.1)$&$\omega_1(0.7)~$& $\beta_m(0.7)$&$\beta_m(0.1)$&
$\omega_{th}~$ \\ \hline
25 & 239  &  3.92& 5.82& 0.96&  1.6 & 0.75 & 0.12& 28 \\ \hline
 8 & 24.5&  1.25 & 5.82& 0.96& 0.76 & 1.6  & 0.25& 3.0 \\ \hline
\end{tabular}
\end{center}
\end{table}

\end{document}